\documentclass[12pt, runningheads]{llncs}
\usepackage{llncsdoc}
\usepackage{amsmath}
\usepackage{amssymb}
\usepackage[all]{xy}           
\usepackage{amssymb}           
\usepackage{fullpage}
\usepackage{hyperref}
\usepackage{eucal}
\usepackage{graphicx}
\usepackage{float}
\usepackage{epsfig}
\usepackage[usenames,dvipsnames]{color}
\usepackage{charter}
\usepackage{color}
\usepackage{comment}
\usepackage{hyperref}
\usepackage{fancyvrb} 
\numberwithin{equation}{section}

\begin{document}

\title{On the influence of social bots in online protests}
\subtitle{Preliminary findings of a Mexican case study}

\titlerunning{On the influence of social bots in online protests}

\author{P. Su\'arez-Serrato\inst{1}
\and M.E. Roberts\inst{2} \and C. Davis\inst{3} \and F. Menczer\inst{3,4}}

\authorrunning{Su\'arez-Serrato, Roberts, Davis, Menczer}

\institute{Instituto de Matem\'aticas, Universidad Nacional Aut\'onoma de M\'exico, Mexico City \\ {\tt pablo@im.unam.mx} \and Department of Political Science, University of California, San Diego \and Center for Complex Networks and Systems Research\\School of Informatics and Computing, Indiana University, Bloomington \and Indiana University Network Science Institute}

\maketitle

\begin{abstract}
Social bots can affect online communication among humans. We study this phenomenon by focusing on \#YaMeCanse, the most active protest hashtag in the history of Twitter in Mexico. Accounts using the hashtag are classified using the \textit{BotOrNot} bot detection tool. Our preliminary analysis suggests that bots played a critical role in disrupting online communication about the protest movement. 
\end{abstract}

\section{Introduction}
On November 7th, 2014, the Mexican Federal District Attorney, Jos\'e Murillo Karam, cued one of his aides towards the end of a press conference by saying {\it ``Ya me cans\'e"} (I am tired).  The press conference was about the status of the investigation into the disappearance of 43 teachers in training from the rural normal school in Ayotzinapa, Guerrero on September 26th, 2014. This gesture of fatigue catalyzed the largest use of a protest hashtag on Twitter in  Mexico to date. 

\subsection{$\#$YaMeCanse} In 2014 Twitter had just over 7 million users in Mexico. According to Crimson Hexagon, the $\#$YaMeCanse hashtag was used in over 2 million tweets during the month following November 7th, 2014, and a total of about 4.4 million times to this date. Its use peaked on 21st November 2014 with 500 thousand posts.
Figure 1 shows the recorded volume and period of activity for $\#$YaMeCanse and some related hashtags. It was claimed during this period that Twitter was being gamed, flooded by accounts that mainly tweeted this hashtag repeatedly, in an attempt to make it more difficult for human users to communicate and find each other using it. As an innovative response, the human users of the hashtag switched to using the $\#$YaMeCanse2 hashtag. As the alleged spamming accounts moved too, the human users subsequently iterated this evasion strategy using $\#$YaMeCanse3, $\#$YaMeCanse4, and so on. This continued with strong use through $\#$YaMeCanse25.

\begin{figure}[b!]
\includegraphics[width=\textwidth]{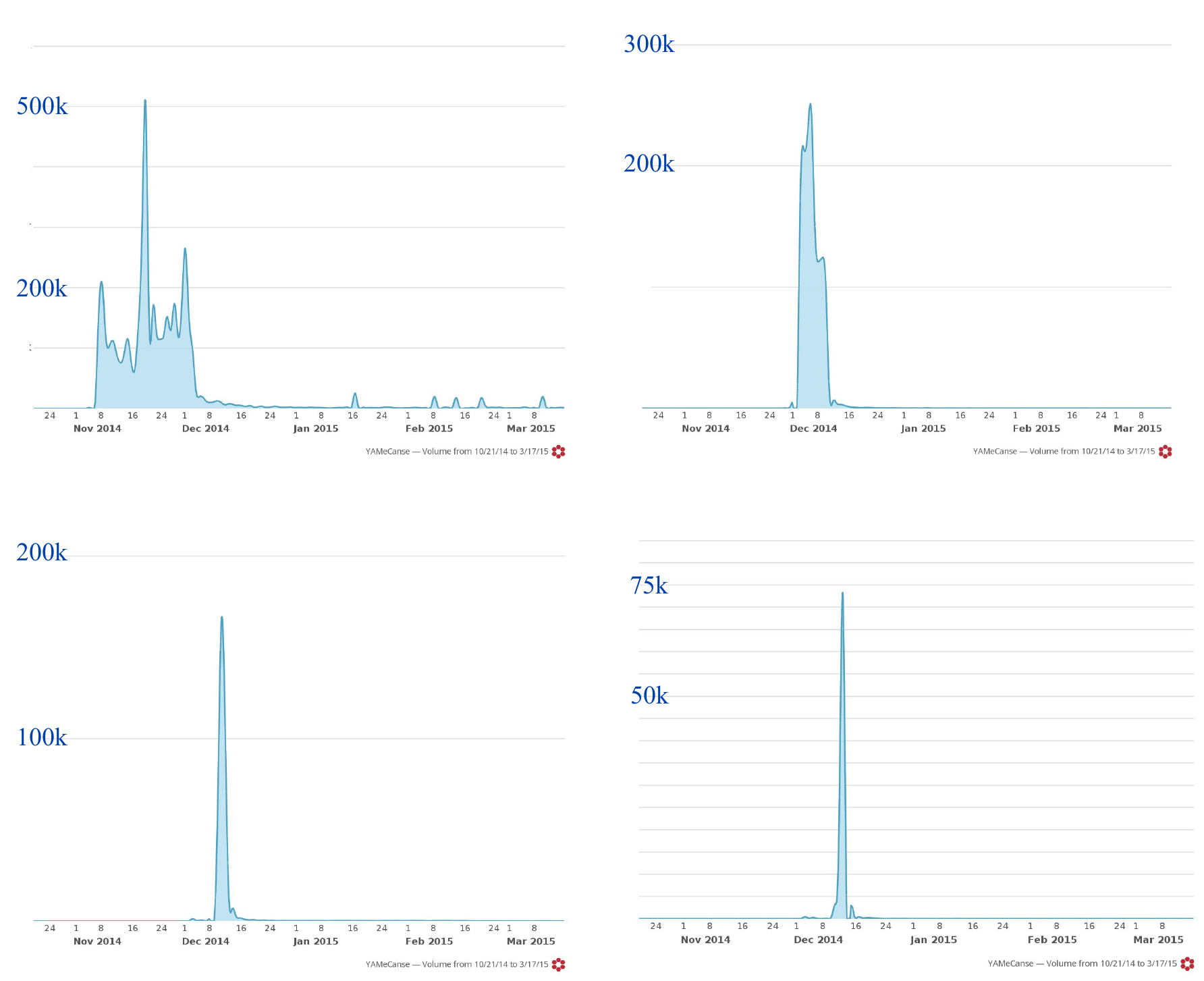}
\caption{ Recorded volume and period of activity for four hashtags: $\#$YaMeCanse (upper left) had strongest use between November 7th and December 14th, 2014;  $\#$YaMeCanse2 (upper right) had strongest use between December 1st and December 16th; $\#$YaMeCanse3 (lower left) had strongest use between December 8th and December 16th; $\#$YaMeCanse4 (lower right) had strongest use between December 9th and December 16th. These diagrams were obtained through Crimson Hexagon.} 
\end{figure}

\subsection{Social Automation}
As companies and institutions strive to automate service processes, there has been a rise in the use of automated social media accounts (for example, chat bots). While some of these accounts can be benign and helpful, some have been designed and deployed with intentions that are not benevolent. Reports of political protests in Russia being swamped by spam on Twitter \cite{BBC News} and of mass pro-government Twitter campaigns in Turkey \cite{Entwickler} have already appeared. 

A bot account can display various levels of automation and content sophistication. They range from spammers that just repeat a hashtag and include irrelevant characters, like punctuation, to bots that repeat a crafted message over and over using multiple---hundreds or even thousands---accounts. {\it Sybil} accounts attempt to disguise themselves as humans. {\it Cyborg} accounts mix automation and human intervention. For example they can be programmed to post at certain intervals. Twitter accounts of news organizations often fall under this category.  A review of the pervasiveness of social bots and the state of the art in detecting them can be found elsewhere \cite{Rise of Social Bots 2016}. 

{\it BotOrNot} is a general supervised learning method for detecting social bot accounts on Twitter \cite{BoN 2016}. This system exploits over 1,000 features that include user meta-data, social contacts, diffusion networks, content, sentiment, and temporal patterns. Based on evaluation on a large set of labeled accounts, {\it BotOrNot} is reported to have high accuracy in discriminating between human and bot accounts, as measured by the Area Under the ROC Curve (AUC 94\%).   
 
It is now understood that the influence of social bots in political discourse, such as a protest, can impede free speech and fracture activist groups \cite{Wooley 2016}. Instances of these outcomes were reported to have also taken place in Mexico in 2015 \cite{Porup 2015}. Recently, Freedom House added pro-government commentators and bots to their analysis of government censorship on the web because these methods can alter the nature and accessibility of the conversation \cite{Freedom House 2015}.

The goal of this research is to establish whether bots interfered with communication between real Twitter users in the context of the $\#$YaMeCanse protest.   We hope to empirically test whether increases in bot activity cause users to lose track of the conversation.  If this is the case, the bots would be functioning as a form of censorship, distracting users from the conversation, similar to online propaganda in other countries like China \cite{KPM 2016}. The ultimate goal of our project
is to quantify bot influence on suppression of communication.

In the preliminary analysis presented here, we use bot identification techniques to verify the involvement of bot accounts in the $\#$YaMeCanse protest.
To this end we present data visualization techniques, using multivariate kernel decomposition estimates and hexagonal bins, to identify regions of potential bots accounts in the phase space obtained by considering pairs of classification probabilities based on different subsets of \textit{BotOrNot} features. These techniques allow us to flag potential bot accounts by focusing on the different outputs produced by \textit{BotOrNot}. This way we can focus on language-independent classifiers, which is important in the present case study because the tweet corpus is in Spanish whereas \textit{BotOrNot} is trained on English content.

 \subsection{Related work}

Twitter bots have been alleged to influence the political discourse surrounding {\it Brexit} in the United Kingdom \cite{Howard 2016}. To evaluate the role of bots, opinions were clustered based on hashtag use. It was found that a very small fraction of most active accounts was responsible for a large fraction of pro-Brexit content: fewer than 2,000 accounts in a collection of 300,000 users (less than 1\%) generated up to 32\% of Twitter traffic about Brexit.

Network decomposition techniques were used to conclude that the success of social protests depends in part on activating a critical periphery \cite{Barbera 2015}. These peripheral participants may be as essential to the communication of the protest message as the most connected and active members. The influence of bots, hindering communication and blocking potential adhesion of new members to a community, could lead to a halt in the movement's growth.

In relation to Mexico, the use of Twitter as a vital communication channel was investigated in the context of urban warfare related to the ongoing war on drugs \cite{Monroy-Hernandez 2012}. Users would tweet the location of conflicts as they erupted, so that their followers could then avoid these violent zones. The authors emphasized that warfare is also a conflict over the control of information, and provided a longitudinal survey of the adoption of Twitter to create safe networks of information.   

The appearance of fake Twitter accounts among the followers of political figures is common, with  20-29\% fake followers in the cases of some prominent people. Accounting for social bots is crucial, for example, in electoral polls and in the identification of influential users on topics of interest. Post sentiment has been studied to separate human from non-human users in Twitter \cite{Dickerson 2014}. A machine learning process was developed, trained on a retrospective analysis of 867 suspended accounts. Political science methods have been used to study Twitter in relation to political issues, with a focus on electoral periods \cite{Gayo-Avello 2013}.

The SentiBot tool employed a combination of graph-theoretic, syntactic, and semantic features to discern between humans, cyborgs, and bots \cite{Chu 2012}. 19 out of the top 25 variables that determine if an account is a bot were found to be related to sentiment. Another classifier exploits natural language processing for social bot detection \cite{Clark 2015}. Such techniques were recently used to investigate the presence and effect of social bots promoting vaporizers and e-cigarettes \cite{Clark 2016}. These are examples of tools that would have to be retrained in order to establish the same results in languages other than English. Porting them to Spanish  would be necessary for application to a corpus of tweets such as the one analyzed in this paper. 
One of our present contributions is the observation that in considering the different classification scores produced by \textit{BotOrNot}, we can use the language-independent features to flag potential bot accounts in Spanish. This technique could potentially be used in other languages as well.

\section{Bot Analysis}

\subsection{Data and Methods}

We had access to datasets of streamed tweets collected by activists using Twitter's streaming API during November and December 2014. The tweets were collected as each hashtag was rising in popularity and only for limited periods, hence the preliminary nature of our analysis. 
The tweets were not stored with all of the Twitter metadata, but the datasets about the first 5 hashtags ($\#$YaMeCanse, $\#$YaMeCanse2, $\ldots$, $\#$YaMeCanse5) have sufficient information for our analysis. In total, these datasets include information from 152,757 tweets. They provide a glimpse into the activity around these hashtags. 
We report on the 28,159 unique accounts that mentioned these five hashtags (Table 1). 

\begin{table}[ht]
\begin{center}
\caption{Total unique users in each dataset}
\begin{tabular}{lr}
    \hline
    Dataset & Unique Users \\
     \hline
    $\#$YaMeCanse & 14756 \\
    $\#$YaMeCanse2 & 1605 \\
    $\#$YaMeCanse3 & 8831 \\
    $\#$YaMeCanse4 & 2530 \\
    $\#$YaMeCanse5 & 437 \\
    \hline
\end{tabular}
\end{center}
\label{tab:unique-users}
\end{table}


\begin{figure}[b!]
\includegraphics[width=\textwidth]{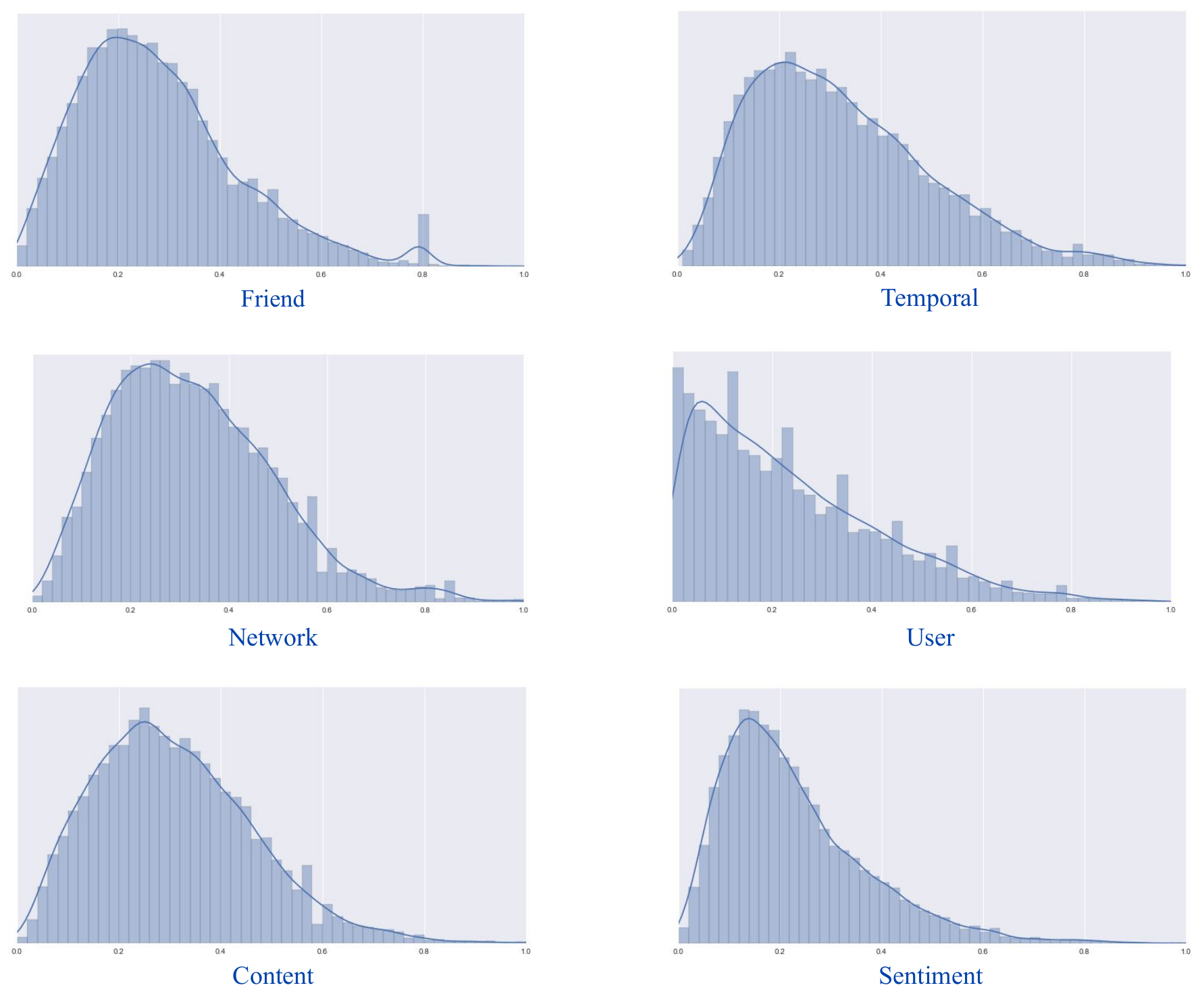}
\caption{\textit{BotOrNot} probability densities for its classifiers using Friend, Temporal, Network, User, Content, and Sentiment features. The distributions are based on 14,756 unique accounts using the $\#$YaMeCanse hashtag between November 26 and 30, 2014.} 
\end{figure}

The user accounts were fed to the \textit{BotOrNot} API \cite{BoN 2016}, taking care to comply with the Twitter API limits. 
%
The \textit{BotOrNot} API includes scores for classifiers trained on subsets of features related to Friends, Network, Time, Content, and Sentiment, as well as an overall score obtained from a classifier trained on all features. Each score can be interpreted as the likelihood that the given account is a bot, according to the features in each of these categories \cite{Rise of Social Bots 2016}.

\subsection{Preliminary Results}

Since most if not all of the tweets are in Spanish, the distributions of Content and Sentiment scores do not register any significant signal pointing to the existence of bots, as illustrated in Figure 2 for the first hashtag, $\#$YaMeCanse. This is not surprising, as the Content and Sentiment classifiers have been trained for English language tweets. The distribution of Friend  scores, however, displays a clear signal: a significant number of accounts have score above 0.8, strongly suggesting the presence of bots. The Friend classifier considers four types of friends (contacts): users retweeting, mentioning, being retweeted, and being mentioned. For each group separately, \textit{BotOrNot} extracts features about number of languages used, local time, account age distribution, popularity distributions, and other user metadata.

\begin{figure}[b!]
\includegraphics[width=\textwidth]{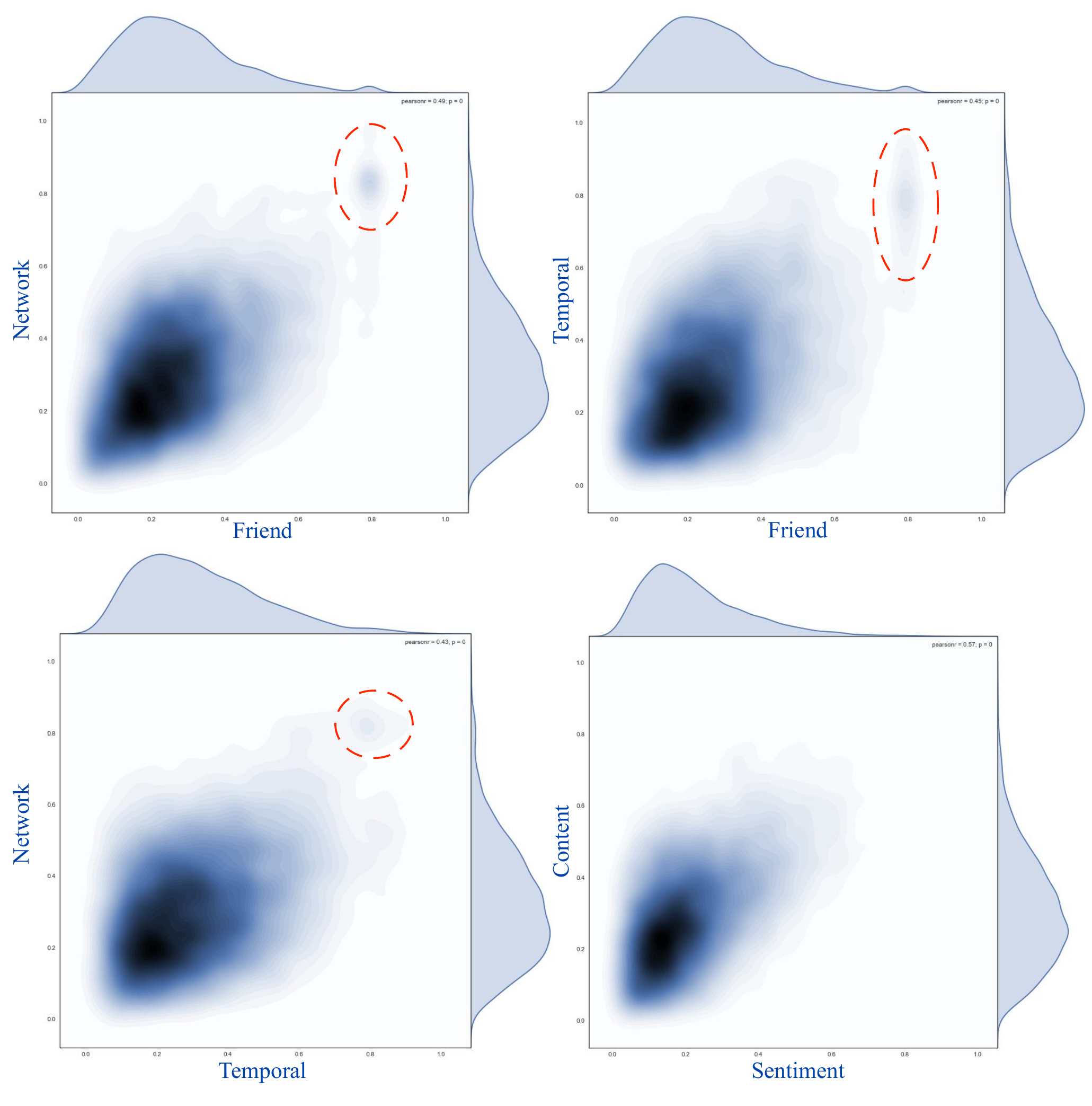}
\caption{ Combined \textit{BotOrNot} probability densities: bi-variate Kernel Density Estimates for pairwise Friend-Network, Friend-Temporal, Temporal-Network and Content-Sentiment classes,  for 14,756 unique accounts using the hashtag $\#$YaMeCanse between November 26 and 30, 2014. } 
\end{figure}

\begin{figure}[b!]
\includegraphics[width=\textwidth]{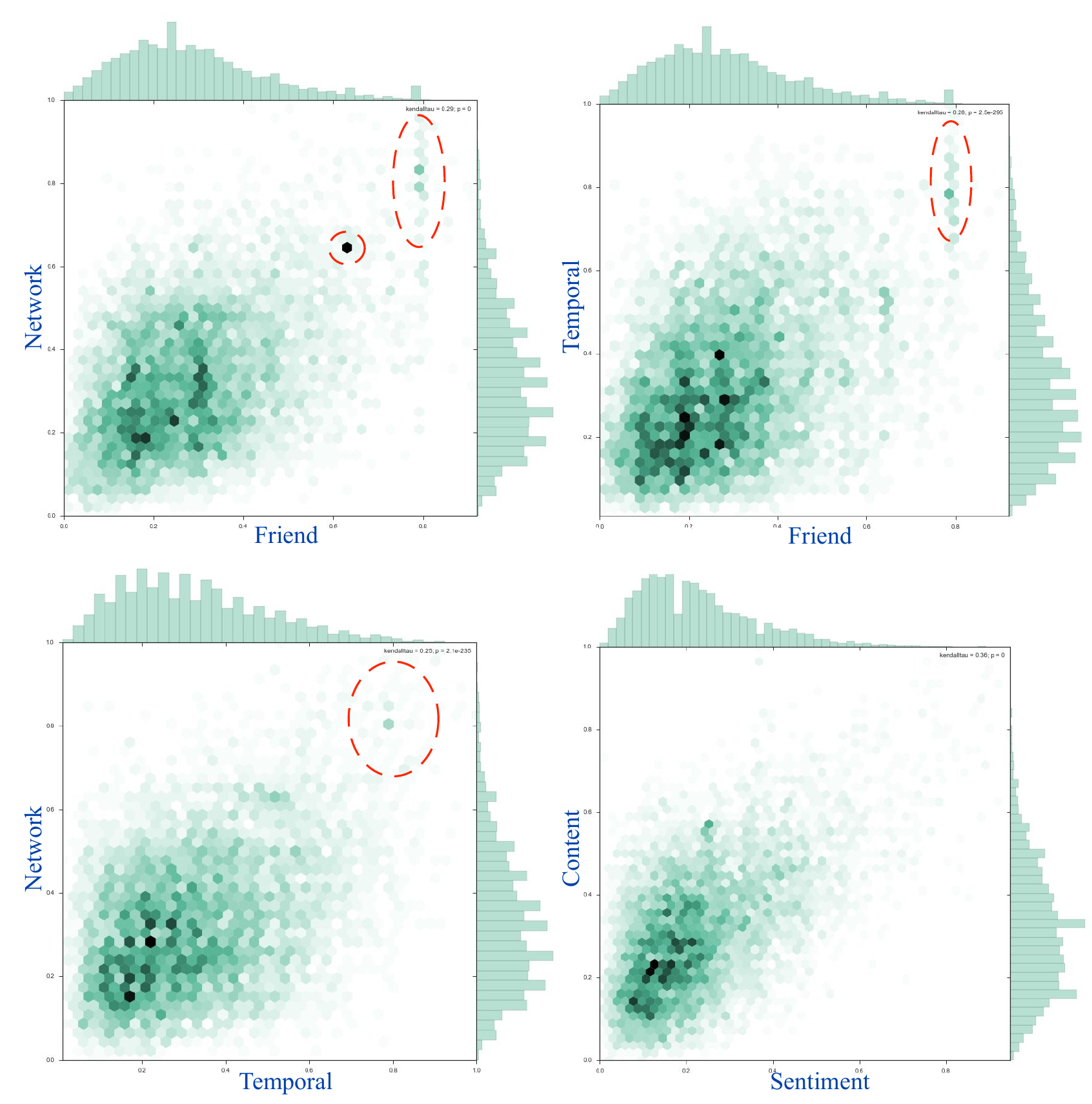}
\caption{ Combined \textit{BotOrNot} probability densities. Bi-variate Hexagonal bins for pairwise Friend-Network, Friend-Temporal, Temporal-Network and Content-Sentiment classes, for 8,831 unique accounts using the hashtag $\#$YaMeCanse3 between December 11 and 13, 2014.} 
\end{figure}

The \textit{BotOrNot} distributions for the Friend, Network, and Temporal classifiers display discernible modes for high bot scores. Therefore we plot these pairwise using bivariate kernel density estimates (Figure 3). The same is done for hexagonal bin plots (Figure 4). 
This analysis is carried out with the Seaborn library \cite{Waskom 2016}. The bot regions lie in the upper-right quadrants above 0.65$\times$0.65 scores. Table 2 summarizes the numbers of bots in these regions. In both figures, we observe clusters of potential bots, grouped away from the human users (marked by dashed lines). In the Content-Sentiment case there is no clear bot cluster. In the Friend-Network hexagonal bin plot, we also notice a second, sharp cluster, suggesting that there may be two distinct types of bots.

\begin{table}[ht]
\begin{center}
\caption{Numbers of likely bot accounts with \textit{BotOrNot} score above 0.65, according to each classifier and in each dataset. For the Temporal classifier we also show the corresponding percentages out of all accounts in each dataset.}
\begin{tabular}{lrrrrrr}
    \hline
Dataset & Temporal & Friend & Network & Content	& Sentiment & User\\
 \hline
    $\#$YaMeCanse &610 (4$\%$)& 340 & 467& 313 & 162 & 365 \\
    $\#$YaMeCanse2 &79  (4$\%$)& 40 & 58 & 44 & 27 & 42 \\
    $\#$YaMeCanse3 &394 (4$\%$)& 259 & 312 & 200 & 110 & 240 \\
    $\#$YaMeCanse4 &130 (5$\%$)& 56 & 83 & 67 & 31 & 63 \\
    $\#$YaMeCanse5 &24  (5$\%$)& 13 & 19 & 9 & 7 & 12 \\
    \hline
\end{tabular}
\end{center}
\label{tab:hi-scorers}
\end{table}

\pagebreak

\section{Conclusions}

It is clear that there was a substantial bot presence affecting the online discussion about the $\#$YaMeCanse protest. In fact, the above estimates are to be considered as lower bounds. While computing bot scores, we ran into many accounts that had been deleted by Twitter (Table 3). Many of these were probably bots. 

A further analysis that incorporates a more representative sample will be carried out in future work, in order to quantify the influence of these bots on the discourse and on the fragmentation among human users of these hashtags. We would also like to look for causal links between volume of tweets produced by bots and shifts between hashtags. To this end we plan to apply an information theoretic approach similar to one previously used for measuring influence in social media \cite{VerSteeg 2012}.

\begin{table}[ht]
\begin{center}
\caption{Deleted accounts found in the datasets.}
\begin{tabular}{lr}
    \hline
       Dataset & Deleted Accounts\\
 \hline
    $\#$YaMeCanse &2084 (14$\%$) \\
    $\#$YaMeCanse2 &235 (14$\%$) \\
    $\#$YaMeCanse3 &1203 (13$\%$) \\
    $\#$YaMeCanse4 &259 (10$\%$) \\
    $\#$YaMeCanse5 &51 (11$\%$) \\
    \hline
\end{tabular}
\end{center}
\label{tab:deleted-accounts}
\end{table}


\subsubsection*{Acknowledgments.} We thank IPAM at UCLA and the organizers of the Cultural Analytics program, where this work was first conceived, for a wonderful working environment and for bringing diverse fields together. PSS is thankful to the Primrose Foundation for coding help. We thank Alberto Escorcia for collecting the tweet data and giving us access to it, and also Twitter for allowing access to data through their APIs.

\end{document}